\documentclass{kluwer}    
\usepackage{epsfig}

\newdisplay{guess}{Conjecture}

\begin{document}                                                                                   
\begin{article}
\begin{opening}         
\title{A unified model for black hole X-ray binary jets ?}
\author{Rob \surname{Fender}}
\runningauthor{Fender, Belloni \& Gallo}
\runningtitle{Black hole X-ray binary jets}
\institute{School of Physics \& Astronomy, University of Southampton, SO17 1BJ, UK}
\author{Tomaso \surname{Belloni}}  
\institute{INAF -- Osservatorio Astronomico di Brera, Via E. Bianchi 46, I-23807 Merate, Italy}
\author{Elena \surname{Gallo}}
\institute{Astronomical Institute 'Anton Pannekoek', University of Amsterdam, Kruislaan 403, 1098 SJ Amsterdam, The Netherlands
}
\date{Feb 17, 2005}

\begin{abstract}
We have recently put forward a 'unified' semi-empirical model for the
coupling between accretion and jet production in galactic black hole
X-ray binaries. In this paper we summarise this model and briefly
discuss relevant considerations which have arisen since its
publication.
\end{abstract}

\end{opening}           
\section{Introduction}

Relativistic jets are a fundamental aspect of accretion onto black
holes on all scales. They can carry away a
large fraction of the available accretion power in collimated flows
which later energise particles in the ambient medium. The removal of
this accretion power and angular momentum must have a dramatic effect
on the overall process of accretion. In their most spectacular form
they are associated with supermassive black holes in active galactic
nuclei (AGN), and with Gamma-Ray Bursts (GRBs), the most powerful and
explosive engines in the Universe respectively. However, parallel
processes, observable on humanly-accessible timescales, are occurring
in the accretion onto black holes and neutron stars in binary systems
within our own galaxy.

We (Fender, Belloni \& Gallo 2004; hereafter FBG04) have recently
published a 'unified' model for the 'disc-jet' coupling in galactic
black hole binaries. In the next few pages we shall quickly summarise
this model.

\section{The model}

In our model we attempt to pin down as accurately as possible the
moment at which the major radio outburst occurred and relate this to
the X-ray state at the time. We subsequently compare this with the
X-ray state corresponding to the lower-luminosity steady jets, to the
evolution of transient outbursts, and to the velocity and power
associated with each 'type' of jet, in order to draw up a framework
for a unified model of black hole X-ray binary jet production.

Several black hole systems are investigated in this paper, and in
addition we compare these with the neutron star systems Cir X-1 and
Sco X-1. The data relating to the radio flares, jet Lorentz factors
(if measured), corresponding X-ray luminosities, estimated distances
and masses, are summarised in table 1 of FBG04. In FBG04 we focus in
particular on the spectral evolution of four black hole binaries, GRS
1915+105, GX 339-4, XTE J1859+226 and XTE J1550-564.  Precise details
of the X-ray data analysis are presented in FBG04. Fig 1 presents the
X-ray (flux and hardness) and radio flux of these four systems around
periods of state transitions. Several key features are apparent from
careful inspection of this figure, in particular, the optically thin
radio outbursts occur around the transition from 'hard VHS (very high
state)/IS (intermediate state)' to 'soft VHS/IS' states, and not at
the transitions to or from the canonical low/hard or high/soft states.

\begin{figure*}
\label{lc}
\centerline{\epsfig{file=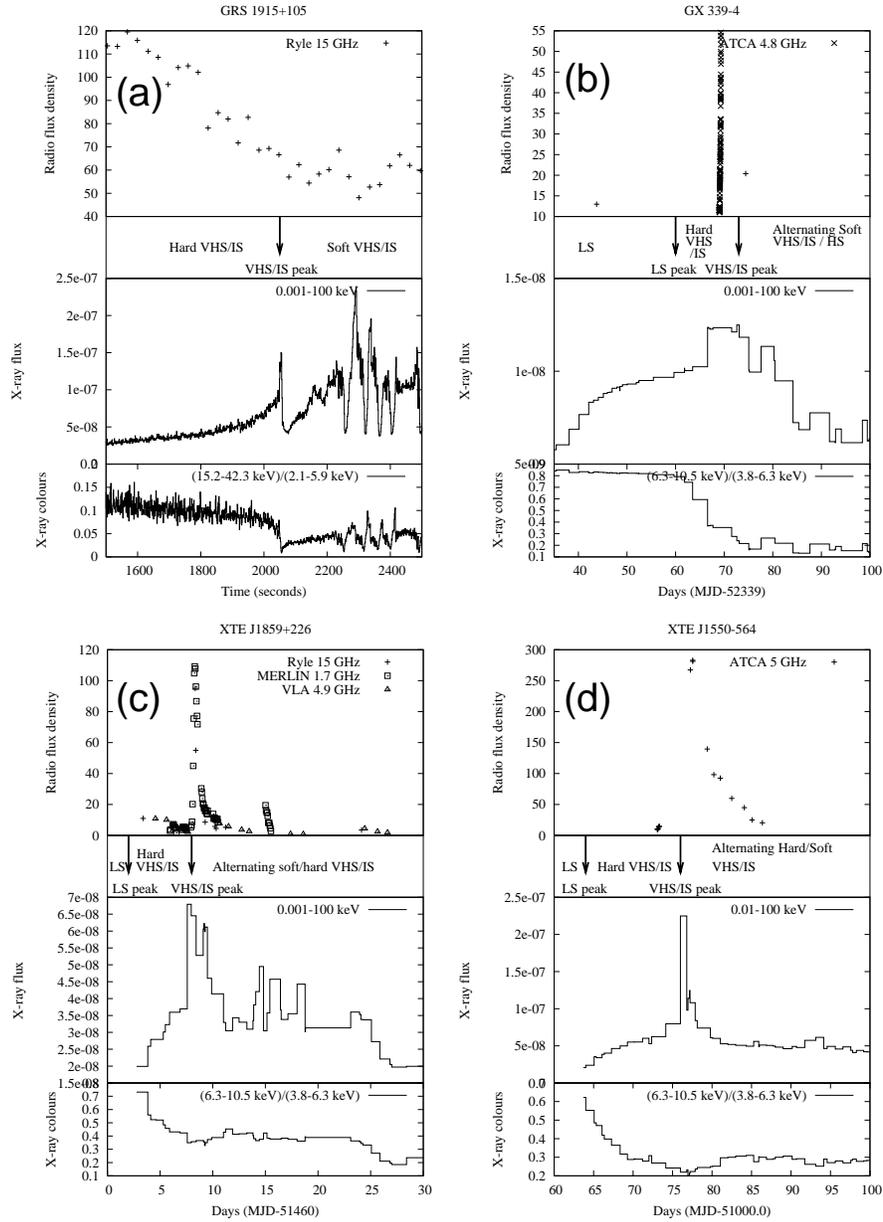, angle=0, width=14cm}}
\caption{Radio and X-ray light curves, X-ray colours and X-ray state
  classifications during periods around transient jet formation, for
  four black hole (candidate) X-ray binaries. In GRS 1915+105 the
  canonical LS or HS are never reached; in GX 339-4, XTE J1859+226 and XTE
  J1550-564 the delay between the canonical LS peak and subsequent
  VHS/IS peak ranges from a few days to two weeks. Nevertheless, in
  all four cases the radio flare occurs at the time of the VHS peak,
  indicating a clear association between this, and not the previous
  LS, and the major ejection. The units of the X-ray flux are erg
  s$^{-1}$ cm$^{-2}$.}
\end{figure*}

\section{Jets as a function of X-ray state: new perspectives}

Based upon the investigation we have performed, we are better able to
associate the characteristics of the radio emission as a function of
X-ray state, and therefore to probe the details of the jet:disc
coupling. While the previously-established pattern of:

\begin{itemize}
\item{LS (low state) = steady jet}
\item{HS (high state) = no jet}
\end{itemize}

remains valid, additional information has clearly come to light about
the details of jet formation in the VHS/IS during transient outbursts.

\subsection{Behaviour of the jet in the 'hard VHS/IS'}

It was previously established that the canonical low/hard state was
associated with a steady jet, the emission from which followed a
'universal' correlation in the $L_{\rm X}$:$L_{\rm radio}$ plane
(e.g. Fender 2001; Corbel et al. 2003; Gallo, Fender \& Pooley
2003). Furthermore the canonical 'high/soft' state was associated with
a dramatic reduction in the radio emission (Tanabaum et al. et
al. 1972; Fender et al. 1999; Gallo, Fender \& Pooley 2003). The study
presented in FBG04 has further revealed that almost until the point of
the major ejection, after the spectrum has started softening (in the
'hard VHS/IS' state) the steady radio jet stays 'on'. 

However, following the persistence of the steady LS-like radio emission into
the hard VHS/IS, the data do indicate that a {\em change} in the radio emission
does occur prior to the radio flare. In brief, it appears that the
radio emission starts to become more variable, with a peaked or (more)
optically thin spectrum shortly before the radio flare. At present we
do not have a clear picture of what is going on during this phase, but
it hints that the major ejection episode is already inevitable some
days in advance of its observational signature.  See FBG04 for more
details.

\begin{figure}
\label{allthree}
\centerline{\epsfig{file=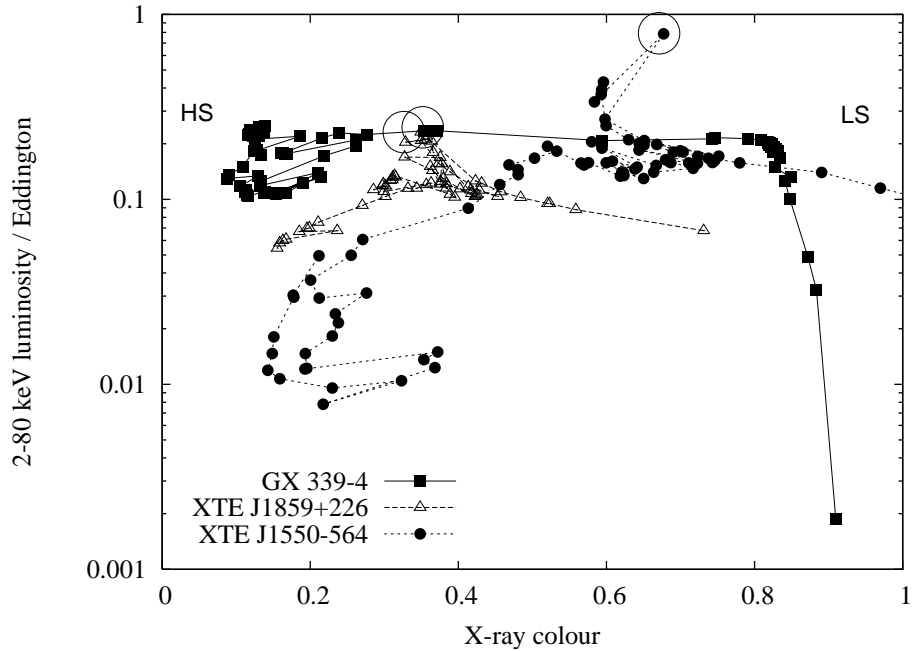, angle=270, width=14cm}}
\caption{Combined X-ray hardness-luminosity diagram (HID) for GX 339-4,
  XTE J1859+226 and XTE J1550-564. The X-ray fluxes plotted in Fig
  1(b--d) have been scaled to Eddington-ratioed luminosities; all the
  sources move approximately anti-clockwise in the diagram. Note that ejections in
  GX 339-4 and XTE J1859+226 occur at almost exactly the same colour
  and X-ray luminosity. Most of the data points correspond to varying
  degrees of the VHS/IS, and not the canonical LS (to the right) or HS
  (to the left).  }
\end{figure}

\subsection{Association of the outburst with the soft VHS/IS peak}

The clearest observational fact to be gleaned from Fig 1 is that the
major optically thin radio outbursts occur in a transition from the
'soft VHS/IS' to the hard 'VHS/IS'. Fig 2 indicates the point of radio
outburst in Hardness -- Luminosity plane.  This aspect is
discussed in considerably more detail in FBG04 but it suffices to note
here that to our knowledge that there are no exceptions to this
pattern of behaviour in any black hole X-ray binary.  It is important
to note that very similar conclusions about the behaviour of the radio
emission in the VHS/IS were drawn by Corbel et al. (2004).

\section{Increasing jet velocity in outburst ?}

In Fig 3 we plot estimated limits on the Lorentz factors
of jets from a handful of X-ray binary systems as a function of X-ray
luminosity at the point of the jet launch. It is important to realise
that the lower-left point is an upper limit on the mean Lorentz factor
of jets in the low/hard state (Gallo, Fender \& Pooley 2003; but see
also Heinz \& Merloni 2004), and that all the other point are lower
limits. The key point is that we have seen highly relativistic motions
from the transient jets associated with outbursts, but not with the
steady low/hard state jets. Thus, the data seem to support the idea
that the powerful jets produced at the transition from the 'hard
VHS/IS' to the 'soft VHS/IS' are at a higher velocity than those which
preceded them. This leads naturally to the likelihood of internal
shocks in the jet, as we shall discuss below.

\begin{figure}
\label{figvelx}
\centerline{\epsfig{file=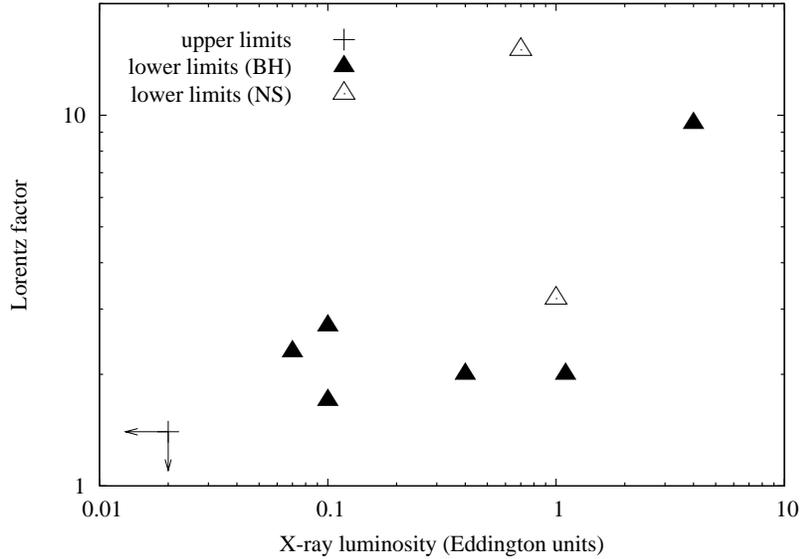, angle=270, width=11cm}}
\caption{Limits on jet Lorentz factors as a function of the estimated
  bolometric X-ray luminosity at the time of jet launch. The arrows in
  the lower left of the figure indicate the condition that jets in the
  general LS state have $v \leq 0.8c$. The rest of the symbols
  are lower limits only to the Lorentz factors from individual black
  hole (filled symbols) and neutron star (open symbols) X-ray
  binaries; the data are listed in table 1.}
\end{figure}

\section{Radio emission and jet power}

It is also crucial to estimate the jet power as a function of X-ray
luminosity / state. In the following we present simplified expressions
for the power in both optically thick and optically thin jets, in
Eddington units, as a function of observable radio and X-ray emission.

\subsection{The low/hard state optically thick jet}

In Fender, Gallo \& Jonker (2003) it was argued that the total jet power
$L_{\rm J}$, in the absence of significant advection, was related to
the accretion luminosity $L_{\rm X}$ as follows:

\[
L_{\rm J} = A_{\rm steady} L_{\rm X}^{0.5}
\]

where $A_{\rm steady} \geq 6 \times 10^{-3}$ (the normalisation is
referred to simply as $A$ in Fender, Gallo \& Jonker 2003).

Studies of the rapid variability from the 'hard' transient XTE
J1118+480, which remained in the LS throughout its outburst, have
supported the idea that the optical emission may originate in an
outflow and not reprocessed emission from the disc (Merloni, di Matteo
\& Fabian 2000; Kanbach et al. 2001; Spruit \& Kanbach 2002; Malzac et
al. 2003). Detailed modelling of the correlated variability by Malzac,
Merloni \& Fabian (2004) has resulted in a normalisation of the
jet/outflow power which corresponds to $A_{\rm steady} \sim 0.3$ in the above
formalisation, which would imply that all LS sources are
jet-dominated. For now we shall take this as the largest likely value
of $A_{\rm steady}$ (see also Yuan, Cui \& Narayan 2004 who estimate a
value for the radiative efficiency for the jet in XTE J1118+480 which
lies between the lower limit of Fender, Gallo \& Jonker 2003 and the
estimate of Malzac, Merloni \& Fabian 2004).

\subsection{The optically thin jets}

The power associated with the production of optically thin jets can be
calculated from the peak luminosity and rise time of the event,
adapting the minimum energy arguments of Burbidge (1959). We
furthermore argue that an additional correction factor of 50 is
applicable to compensate for bulk relativistic motion (see FBG04 for
details).

A best-fit power-law to the data for the transient events is of the form

\[
L_{\rm jet} = A_{\rm trans} L_{\rm X}^{0.5 \pm 0.2}
\]

where the fitted value is $A_{\rm trans} = (0.4 \pm 0.1)$, within
uncertainties the same index as inferred for the steady jets. Note that
since for the transient jets $L_{\rm X} \sim 1$ (in Eddington units)
this indicates near equipartition of $L_{\rm X}$ and $L_{\rm J}$
around the time of such events.

\section{Internal shocks}

The arguments given above clearly indicate that as the X-ray
luminosity of the accreting source increases, then so does the
velocity of the outflow (although whether this is in the form of a
step, or other functional form, is as yet unclear). Since most,
probably all, outbursting sources have followed a path in which they
have become monotonically brighter in a hard state before making a
transition to a soft state, this tells us that a shock should form in
the previously-generated `steady' jet as the faster-moving VHS/IS jet
catches up and interacts with it. This internal shock is therefore a
natural origin for the optically thin events observed at the beginning
of X-ray transient outbursts. Internal shocks have previously been
proposed for AGN (e.g. Rees 1978; Marscher \& Gear 1985; Ghisellini
1999; Spada et al. 2001) and gamma-ray bursts (GRBs) (e.g. Rees \&
Meszaros 1994; van Paradijs, Kouveliotou \& Wijers 2000 and references
therein).  Indeed in the context of X-ray binaries an internal-shock
scenario has already been discussed previously for GRS 1915+105 by
Kaiser, Sunyaev \& Spruit(2000), Vadawale et al. (2003) and Turler et
al. (2004), and their ideas have significantly inspired this work.  In
the context of the changes in Lorentz factor estimated here, internal
shock efficiencies as high as 30\% may be possible, although lower
efficiencies seem more likely.

Internal shocks at relatively large distances from the base of the jet
are a natural explanation for why the emission in such outburst is
optically thin, unlike the steady self-absorbed jet which preceded it.
Also, as discussed in Vadawale et al. (2003) the strength of the shock is
likely to be related to the amount of material lying in the path of
the faster 'VHS/IS' jet. They discussed this in the context of GRS
1915+105, where the strength of 'post-plateau jets' (Klein-Wolt et
al. 2002) is shown to be correlated with the total X-ray fluence of
the preceding 'plateau' (which was presumably a phase of slower jet
production). Generalising this phenomenon to other X-ray transients,
it provides a natural explanation for why, although there are often
multiple radio flaring events, the first is invariably the strongest.

\begin{figure*}
\label{toymodel}
\centerline{\epsfig{file=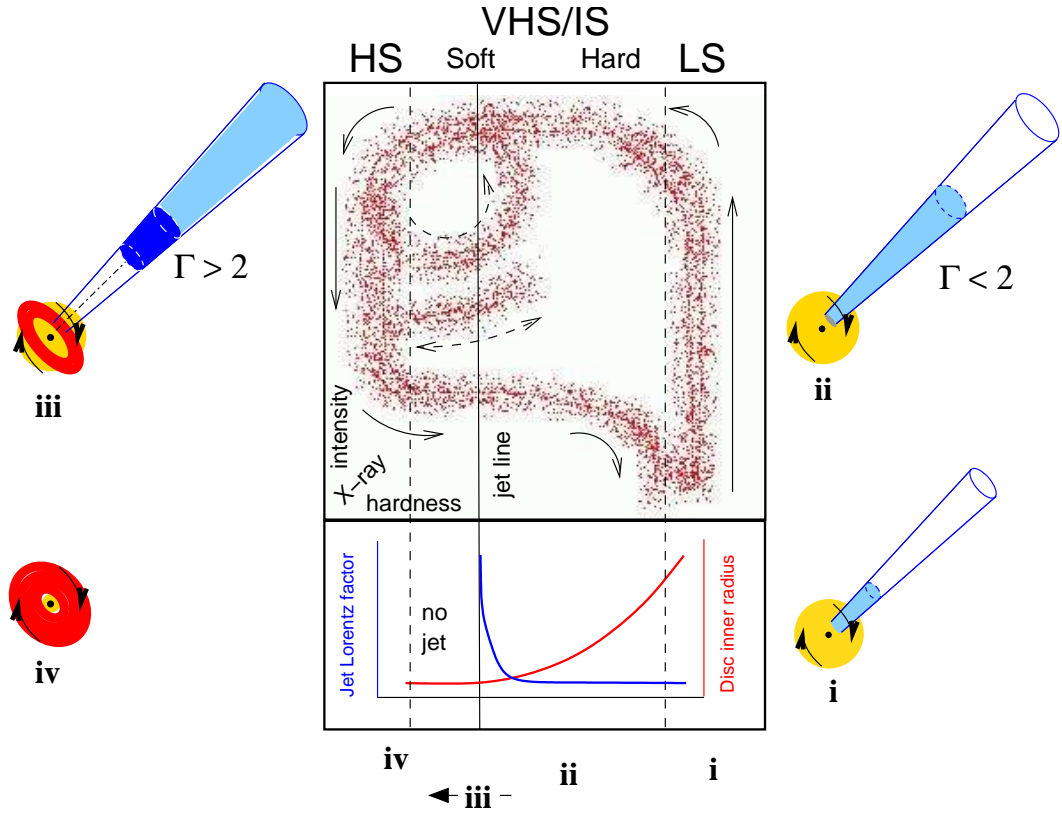, angle=0, width=14cm}}
\caption{A schematic of our simplified model for the jet-disc coupling
  in black hole binaries. The central box panel represents an X-ray
  hardness-intensity diagram (HID); 'HS' indicates the `high/soft
  state', 'VHS/IS' indicates the 'very high/intermediate state' and
  'LS' the 'low/hard state'. In this diagram, X-ray hardness increases
  to the right and intensity upwards. The lower panel indicates the
  variation of the bulk Lorentz factor of the outflow with hardness --
  in the LS and hard-VHS/IS the jet is steady with an almost constant
  bulk Lorentz factor $\Gamma < 2$, progressing from state {\bf i} to
  state {\bf ii} as the luminosity increases. At some point -- usually
  corresponding to the peak of the VHS/IS -- $\Gamma$ increases
  rapidly producing an internal shock in the outflow ({\bf iii})
  followed in general by cessation of jet production in a
  disc-dominated HS ({\bf iv}). At this stage fading optically thin
  radio emission is only associated with a jet/shock which is now
  physically decoupled from the central engine.  As a result the solid
  arrows indicate the track of a simple X-ray transient outburst with
  a single optically thin jet production episode. The dashed loop and
  dotted track indicate the paths that GRS 1915+105 and some other
  transients take in repeatedly hardening and then crossing zone {\bf
  iii} -- the 'jet line' -- from left to right, producing further optically thin radio
  outbursts. Sketches around the outside illustrate our concept of the
  relative contributions of jet (blue), 'corona' (yellow) and
  accretion disc (red) at these different stages.}
\end{figure*}

\section{Towards a unified model}

Based upon the key generic observational details assembled above, we
have attempted to construct a unified, semi-quantitative, model for
the disc-jet coupling in black hole X-ray binaries. A simplified
version of the model specific to GRS 1915+105 has been presented in
Fender \& Belloni (2004).  The model is summarised in Fig 4, which we
describe in detail below. The diagram consists of a schematic X-ray
hardness-intensity diagram (HID) above a schematic indicating the bulk
Lorentz factor of the jet and inner accretion disc radius as a
function of X-ray hardness. The four sketches around the outside of
the schematics indicate our suggestions as to the state of the source
at the various phases {\bf i}--{\bf iv}. The path of a typical X-ray
transient is as indicated by the solid arrows.

\begin{itemize}
\item{Phase {\bf i}: Sources are in the low-luminosity LS, producing a
  steady jet whose power correlates as $L_{\rm jet} \propto L_{\rm
  X}^{0.5}$ (ignoring any mass term). This phase probably extends down
  to very low luminosities ('quiescence')}.
\item{Phase {\bf ii}: The motion in the HID, for a typical outburst,
  has been nearly vertical. There is a peak in the LS after which the
  motion in the HID becomes more horizontal (to the left) and the
  source moves into the 'hard' VHS/IS. Despite this softening of the
  X-ray spectrum the steady jet persists, with a very similar
  coupling, quantitatively, to that seen in the LS.}
\item{Phase {\bf iii}: The source approaches the 'jet line' (the
  solid vertical line in the schematic HID) in the HID between jet-producing
  and jet-free states. As the boundary is approached the jet
  properties change, most notably its velocity. The final, most
  powerful, jet, has the highest Lorentz factor, causing the
  propagation of an internal shock through the slower-moving outflow
  in front of it.}
\item{Phase {\bf iv}: The source is in the 'soft' VHS/IS or the
  canonical HS, and no jet is produced. For a while following the peak
  of phase iii fading optically thin emission is observed from the
  optically thin shock.}
\end{itemize}

Following phase {\bf iv}, most sources drop in intensity in the
canonical HS until a (horizontal) transition back, via the VHS/IS, to
the LS. Some sources will make repeated excursions, such as the loops
and branches indicated with dashed lines in Fig 4, back across the jet
line, However, with the exception of GRS 1915+105, the number of such
excursions is generally $\leq 10$.  When crossing the jet line from
right to left, the jet is re-activated but there is (generally) no
slower-moving jet in front of it for a shock to be formed; only motion
from left to right produces an optically thin flare (this is a
prediction). Subsequently the motion back towards quiescence is almost
vertically downwards in the HID.

The model as outlined above has many similarities with the scenarios
described by Meier (1999, 2001, 2003) who has approached the problem
from a more theoretical point of view. Meier (2001) has suggested that
in low-luminosity states the jet is powered by a modification of the
Blandford \& Payne ('BP') (1982) mechanism taking into account frame-dragging
near a rotating black hole (Punsly \& Coroniti 1990). This 'BP/PC
mechanism' can extract black hole spin by the coupling of magnetic
field lines extending from within the ergosphere to outside of
it. Meier (2001) further suggests that during phases of very high
accretion the Blandford \& Znajek ('BZ') (1977) mechanism may work
briefly. This may be associated with a 'spine jet' which is
considerably more relativistic than the 'sheath jet' produced by the
BP/PC mechanism. Note that the power of the jets as given in Meier
(2001, 2003) is about linearly proportional to the accretion rate; in
the formulation of Fender, Gallo \& Jonker (2003) this corresponds to
the 'jet dominated state' (see also Falcke, Kording \& Markoff 2004).

We can revisit the scenarios of Meier in the light of our compilation
of observational results and steps toward a unified model. In the
faint LS (phase {\bf i} in Fig 4) is the jet formed by the BP or BP/PC
mechanisms ?  Given that the jet may be formed at relatively large
distances from the black hole in such states, there may not be any
significant influence of the black hole spin on the jet formation
process. However, it is also likely that in such states the
jet-formation process is not occurring within thin discs, as is the
basis of the BP mechanism, but rather in a geometrically thick flow
(see also e.g. Blandford \& Begelman 1999; Meier 2001; Merloni \&
Fabian 2002)

As the accretion rate increases the power of this disc-jet will
increase and the geometrically thin accretion disc will propagate
inwards. During this phase the jet formation process may migrate from
BP$\rightarrow$BP/PC. However, the suggestion that the most
relativistic jets are formed by the BZ process seems at odds with the
observation of significantly relativistic outflows from two neutron
stars systems (Fomalont et al. 2001a,b; Fender et al. 2004).  In a
related work, the results of Yu, van der Klis \& Fender (2004)
indicate that the subsequent evolution of X-ray transient outbursts is
approximately determined {\em before} the soft VHS/IS peak, in both
neutron star and black hole systems. This suggests that already by
the time of the LS peak we can estimated the size of the ejection even
which is to follow, and is a further indication that the study of neutron
stars will shed important light on the physics of jet formation in
black hole systems.

\section{Summary of the model}

We have examined the observational properties of the jets associated
with black hole X-ray binary systems. The key observations can be
summarised as:

\begin{enumerate}
\item{{\bf The radio:X-ray coupling:} we have established that the
  steady radio emission associated with the canonical LS persists
  beyond the softening of the X-ray spectrum in the 'hard' VHS/IS. At
  the end of the transtion from 'hard' to 'soft' VHS/IS, usually
  associated with a local maximum in the X-ray light curve, a
  transient radio outburst occurs. The radio emission is subsequently
  suppressed until the source X-ray spectrum hardens once more. Some
  source may repeatedly make the transition from 'hard' to 'soft'
  VHS/IS and back again, undergoing repeated episodes of steady and
  transient jet formation. }
\item{{\bf Jet velocities:} we have argued that the
  measured velocities for the transient jets, being
  relativistic with $\Gamma \geq 2$ are significantly larger than those
  of the steady jets in the LS, which probably have $\Gamma \leq 1.4$.}
\item{{\bf Jet power:} we have furthermore established that our best
 estimates of the power associated with the transient jets are
 compatible with extrapolations of the functions used to estimate the
 power in the LS (albeit with a relatively large normalisation).
}
\end{enumerate}

Essentially equivalent conclusions about the radio:X-ray coupling have
been drawn by Corbel et al. (2004). Putting these observational aspects
together we have arrived at a semi-quantitative model for jet
production in black hole XRBs. We argue that for X-ray spectra harder
than some value (which may be universal or vary slightly from source
to source) a steady jet is produced. The power of this jet correlates
in a non-linear way (approximately given as $L_{\rm J} \propto L_{\rm
X}^{0.5}$) with the X-ray luminosity. As the X-ray luminosity
increases above $\sim 1$\% of the Eddington rate the X-ray spectrum
begins to soften. Physically this probably corresponds to the heating
of the inner edge of the accretion disc as it propagates inwards with
increasing accretion rate. Initially the jet production is not
affected. As the disc progresses inwards the jet velocity
increases. As it moves through the last few gravitational radii before
the ISCO, the Lorentz factor of the jet rises sharply, before the jet
is suppressed in a soft disc-dominated state. The rapid increase in
jet velocity in the final moments of its existence results in a
powerful, optically thin, internal shock in the previously existing
slower moving outflow.

The inner disc may subsequently recede, in which case a steady jet is
reformed, but with decreasing velocity and therefore no internal
shocks. If the disc once more moves inwards and reaches the 'fast jet'
zone, then once more an internal shock is formed. In fact while jets
are generally considered as 'symptoms' of the underlying accretion
flow, we consider it possible that the reverse may be true. For
example, it may be the 'growth' of the steady jet (via e.g. build up
of magnetic field near the ISCO / black hole) which results in the
hardening of the X-ray spectrum, perhaps via pressure it exerts on the
disc to push it back, or simply via Comptonisation of the inner disc
as it spreads (for further discussions see e.g. Nandi et al. 2001;
Tagger et al. 2004).

In the context of the nature and classification of black hole
'states', these states, whether 'classical' or as redefined by
McClintock \& Remillard (2004) do not have a one-to-one relation with
the radio properties of the source. It seems that as far as the jet is
concerned, it is 'on' -- albeit with a varying velocity -- if the disc
does not reach 'all the way in', which probably means as far as the
ISCO. The dividing 'jet line' (Fig 7) HID, may also correspond, at
least approximately, to a singular switch in X-ray timing properties
(Belloni 2004; Belloni et al. 2004; Homan \& Belloni 2004; see also
the discussion in McClintock \& Remillard 2004) and may be the single
most important transition in the accretion process. Further study of
the uniqueness of the spectral and variability properties of sources
at this transition point should be undertaken to test and refine our
model.

Finally, given that Merloni, Heinz \& di Matteo (2003) and Falcke, K\"ording \&
Markoff (2004) (see also Heinz \& Sunyaev 2003; Maccarone, Gallo \&
Fender 2003) have recently demonstrated quantitatively the scaling of
radio:X-ray coupling across a range of $\geq 10^7$ in black hole mass,
it is obviously of great interest to see if the model we are working
towards for the coupling of accretion and jet formation in black hole
binaries may also be applied to AGN. In addition, detailed modelling
of the internal shock scenario is required to see if the coupling, as
outlined above, really could allow us to predict radio light curves
from X-ray, and vice versa. These two areas should be the next steps
forward.

\section{And afterwards....}

The model outlined above was published in December 2004, five months
ago at the time of writing. Since this time we have not discovered, or
been made aware of, any major flaws. There is of course still plenty
of time, and exceptions to the patterns discussed will certainly be
found (and we encourage people to be critical!).

In particular Homan \& Belloni (2005 $\rightarrow$ these proceedings)
have demonstrated very clearly that the neat shape sketched out for
the evolution of black hole outbursts in the HID varies from source to
source, and indeed between outbursts for the same source. Clearly the
sketch should be considered as an idealisation; nevertheless the
overall properties of the pattern in the HID are consistent with our
simple sketch.

Furthermore, both Homan \& Belloni and Remillard (2005) have noted that the
X-ray variability behaviour of the black holes, in particular of
different types of QPOs, may fit empirically into the model. This is
clearly an exciting next step.

Both of these points indicate ways in which the model can be
significantly refined, especially in the context of adding the timing
properties. 

Finally, several people have discussed with us possible applications
of the model to AGN. In particular we would like to note the
implication, first spotted by Chris Simpson, that if AGN follow the
same patterns of outburst there should be some sources with
radio-quiet cores and yet 'relic' radio lobes.

\section*{Acknowledgements}

RPF would like to thank many people for useful discussions related to
the ideas presented here, including Catherine Brocksopp, Annalisa
Celotti, Stephane Corbel, Peter Jonker, Marc Klein-Wolt,
Tom Maccarone, Dave Meier, Simone Migliari, Jon Miller, Felix 
Mirabel, Chris Simpson and {\em especially} Jeroen Homan.

\end{article}
\end{document}